\documentclass[conference]{IEEEtran}
\IEEEoverridecommandlockouts

\usepackage{cite}
\usepackage{amsmath,amssymb,amsfonts}
\usepackage{algorithmic}
\usepackage{graphicx}
\usepackage{textcomp}
\usepackage{xcolor}
\usepackage{booktabs}
\usepackage{subcaption}
\usepackage{float}
\def\BibTeX{{\rm B\kern-.05em{\sc i\kern-.025em b}\kern-.08em
    T\kern-.1667em\lower.7ex\hbox{E}\kern-.125emX}}
\begin{document}

\title{A Data-Free Membership Inference Attack on Federated Learning in Hardware Assurance\\
}


\newif\ifshowauthors
\showauthorstrue 



\ifshowauthors
\author{\IEEEauthorblockN{Gijung Lee, Wavid Bowman, Olivia P. Dizon-Paradis, Reiner N. Dizon-Paradis, \\Ronald Wilson, Damon L. Woodard, Domenic Forte}
\IEEEauthorblockA{Florida Institute of National Security, University of Florida, Gainesville, Florida, USA\\
Email: \{lee.gijung, wavid.bowman, paradiso, reinerdizon, ronaldwilson\}@ufl.edu, \{dwoodard, dforte\}@ece.ufl.edu}}

\else 
\author{
\IEEEauthorblockN{
Author names have been removed for blind review.\\}}
\fi

\maketitle

\begin{abstract}
Federated Learning (FL) is an emerging solution to the data scarcity problem for training deep learning models in hardware assurance. While FL is designed to enhance privacy by not sharing raw data, it remains vulnerable to Membership Inference Attacks (MIAs) that can leak sensitive intellectual property (IP). Traditional MIAs are often impractical in this domain because they require access to auxiliary datasets that can match the unique statistical properties of private data. This paper introduces a novel, data-free MIA targeting image segmentation models in FL for hardware assurance. Our methodology leverages Standard Cell Library Layouts (SCLLs) as priors to guide a gradient inversion attack, allowing an adversary to reconstruct images from a client's intercepted model update without needing any private data. We demonstrate that, by analyzing the reconstruction fidelity, an adversary can infer sensitive hardware characteristics, successfully distinguishing between circuit layers (e.g., metal vs. diffusion) and technology nodes (e.g., 32nm vs. 90nm). Our findings reveal that a novel loss term can conditionally amplify the attack's effectiveness by overcoming evaluation bottlenecks for structurally complex data. This work underscores a significant IP risk, challenging the assumption that FL provides inherent privacy guarantees and proving that severe information leakage can occur even without access to domain-specific datasets.
\end{abstract}


\section{Introduction}\label{sec:introduction}
Semiconductor manufacturers increasingly depend on third‑party IP and external vendors across many stages of the integrated circuit (IC) supply chain \cite{sharma2021new}. The globalized IC supply chain contains serious security threats, including hardware Trojans, IP piracy, IC counterfeiting, and IC overproduction \cite{rostami2014primer}. As a result, hardware assurance has emerged as a core national priority, as emphasized in the CHIPS Act \cite{senate}. To counter these threats, deep learning (DL) technology is being utilized as a key tool for automating security analysis. AI models perform sophisticated tasks such as identifying Trojans that are difficult to detect with conventional methods, as well as chip reverse-engineering and defect analysis from scanning electron microscopy (SEM) images. However, these advanced AI models have an inherent limitation: they require a vast amount of specialized data for their training process. The data needed for model training can only be obtained through the high-cost, slow process of de-layering a chip and capturing SEM images, which creates a serious bottleneck for technological advancement. Consequently, it is challenging for a single organization to acquire enough data to train the advanced models. 

Federated Learning (FL)---technique that allows multiple organizations to collaboratively train a single powerful AI model without directly sharing their sensitive raw data---has emerged as a promising approach to solve this data scarcity problem. Although FL is structured to protect user privacy by avoiding direct raw data exposure, it is still vulnerable to serious privacy compromises. A primary example is the Membership Inference Attack (MIA) \cite{shokri2017membership, pyrgelis2017knock}, a foundational privacy attack where an adversary seeks to ascertain whether a particular piece of data was used when training the model \cite{bai2024membership}.
Most existing MIA studies in the FL environment \cite{nasr2019comprehensive, 9148790, zari2021efficient, li2023effective} concentrate exclusively on analyzing information from the targeted client's updates, utilizing metrics such as gradient norms, loss values, and gradient differences. Recently, research has emerged \cite{gu2022cs, he2024enhance} that attempts to boost the effectiveness of MIAs by introducing shadow models into the FL environment. While using shadow models can offer attackers extra information, these techniques depend on having a distinct auxiliary dataset for training these models. This assumption might be impractical in typical FL scenarios where the central server lacks access to the private data held by clients \cite{zhu2025fedmia}. 

Our core motivation stems from the distinct challenges of applying FL to hardware assurance. Proprietary datasets, such as collections of IC layouts or SEM images, are highly sensitive intellectual property (IP). Traditional attacks fail because public datasets cannot match the unique statistical properties of private hardware data, which vary significantly across different semiconductor manufacturing processes (e.g., 32nm vs. 90nm nodes). To overcome these limitations, this paper introduces a novel MIA methodology targeting image segmentation models trained via FL in the hardware assurance domain.
Our key insight is that an adversary can use the detailed layout masks from widely accessible standard cell library layouts (SCLLs) as powerful candidate labels to guide a gradient inversion attack (GIA) \cite{zhu2019deep} which reconstructs private training data by optimizing dummy inputs until their gradients match observed target (victim) gradients (e.g., from federated learning), thereby revealing sensitive information. By intercepting a client's update from the FL model, our method attempts to reconstruct the original image from the target's dataset. We hypothesize that data points that were true members of the target's training set will exhibit significantly higher reconstruction fidelity during this process. This allows us to perform a MIA by analyzing the quality of the reconstructed images via a GIA, determining if a specific data point was part of the training set. 

Crucially, in hardware assurance domain, this membership inference extends beyond individual data points, as an adversary can infer the specific hardware metadata (node or layer) characteristic of the client's training data. This leakage of hardware-specific membership information is a critical enabler for physical attacks, as it accelerates reverse engineering \cite{torrance2009state}, facilitates identification of design weaknesses~\cite{courbon2020practical} to launch non-invasive attacks, and streamlines IP piracy \cite{guin2014counterfeit}.
The main contributions of our work are as follows:
\begin{itemize}
    \item A Novel Data-Free Attack: We propose the first MIA methodology targeting FL segmentation models that uses SCLLs to guide a GIA, eliminating the need for any domain-specific private datasets for the attack.
    \item Hardware-Specific Inference: We demonstrate that an adversary can infer specific hardware characteristics of a client's private data, successfully distinguishing between layers (e.g., metal vs. diffusion) and technology nodes (e.g., 32nm vs. 90nm), by measuring the differential reconstruction success using various SCLLs as priors for the GIA.
    \item Reconstruction-Based Membership Detection: We validate that reconstruction fidelity is effective metric for distinguishing members from non-members in the hardware assurance domain.
\end{itemize}

The remainder of this paper reviews background on FL and MIAs (Section~\ref{sec:background}), details our methodology and threat model (Section~\ref{sec:methodology}), presents our results (Section~\ref{sec:results}), discusses our results (Section~\ref{sec:discussion}), suggests future works (Section~\ref{sec:futurework}), and concludes (Section~\ref{sec:conclusion}).
\vspace{-1mm}

\section{Background}\label{sec:background}
In this section, we review the main concepts of FL and explain how MIAs are defined in this setting.

\subsection{Federated Learning (FL)}\label{Federated Learning:background}
FL is technique that allows multiple parties to collaboratively train a model without exchanging their private data \cite{Wen2023federated}. In this approach, instead of moving private data to a central server, the local model is trained on each party's own dataset. Only the resulting model updates, such as parameter changes, are shared to update the global model. The goal of this approach is to enhance data privacy by securing sensitive information, which never leaves its local environment, thereby reducing the risk of data breaches. FL also fosters secure collaboration between different organizations, such as researchers, manufacturers, and intellectual property (IP) vendors. It enables them to jointly develop advanced deep learning (DL) models while protecting their confidential information and proprietary IP. The process of FL involves several key steps.
The FL protocol is typically executed in iterative rounds, with each round following these fundamental steps \cite{torkzadehmahani2022privacy}:
\begin{enumerate}
    \item Selection and Distribution: The process begins when the central server selects a group of parties for a training iteration and distributes the most recent version of the global model to them.
    \item Local Training: Each selected party then trains this model using its own private data. This local training computes a set of local updates (for example, parameter gradients) that represent the insights learned from its unique dataset.
    \item Update Aggregation: After local training is complete, the parties send only their computed updates back to the central server. The server then aggregates these individual contributions, by averaging them, to create a new, more refined global model.
    \item Iteration: This newly improved global model becomes the starting point for the next training round, and the cycle repeats.
\end{enumerate}

FL protocols like Federated Stochastic Gradient Descent (FedSGD) and Federated Averaging (FedAVG) \cite{mcmahan2017communication} differ in their update mechanisms. In FedSGD, clients directly transmit gradients to a central server, allowing an attacker to intercept these gradients to infer the original training data. In contrast, FedAVG requires clients to train a model locally for several epochs and share the updated model weights. While this avoids direct gradient sharing, an attacker can still perform a GIA by first estimating the gradients from the difference between the model weights before and after the local training.

\subsection{Membership Inference Attacks (MIAs)}\label{Membership Inference Attacks:background}

A MIA is a privacy violation where an adversary tries to identify if a particular data record was included in a model's training set. When such an attack succeeds, the confidentiality of that data is compromised. For instance, in the context of hardware assurance, consider a model trained on a specific set of SEM images. An adversary could probe this model by providing it with images of both a 32nm metal layer and a 32nm diffusion layer. If the model classifies the 32nm metal layer with an unusually high confidence score, the adversary can infer that images of this structure were likely part of the training set. Furthermore, an adversary with knowledge of a foundry's standard cell library from a specific Process Design Kit (PDK) could exploit this. By testing the model against various SEM images from that library, the adversary could determine if the model was trained using images from that specific, proprietary PDK. Such attacks represent a significant threat to data confidentiality, as they can lead to the leakage of valuable intellectual property. Pioneering research on MIAs was conducted by \cite{shokri2017membership}, who proposed the shadow training technique. 
The shadow training technique requires an attacker to first use a ``shadow dataset" which contains data with a similar distribution to the target's (victim). With this dataset, the attacker trains a ``shadow model" that mirrors the target's architecture and collects its output prediction vectors. These vectors are subsequently used to train a final attack model capable of determining a data record's membership with an accuracy greater than random guessing.
\begin{figure*}[ht]
\vspace{-2mm}
\centering
\includegraphics[width=0.9\linewidth, height=0.4\linewidth]{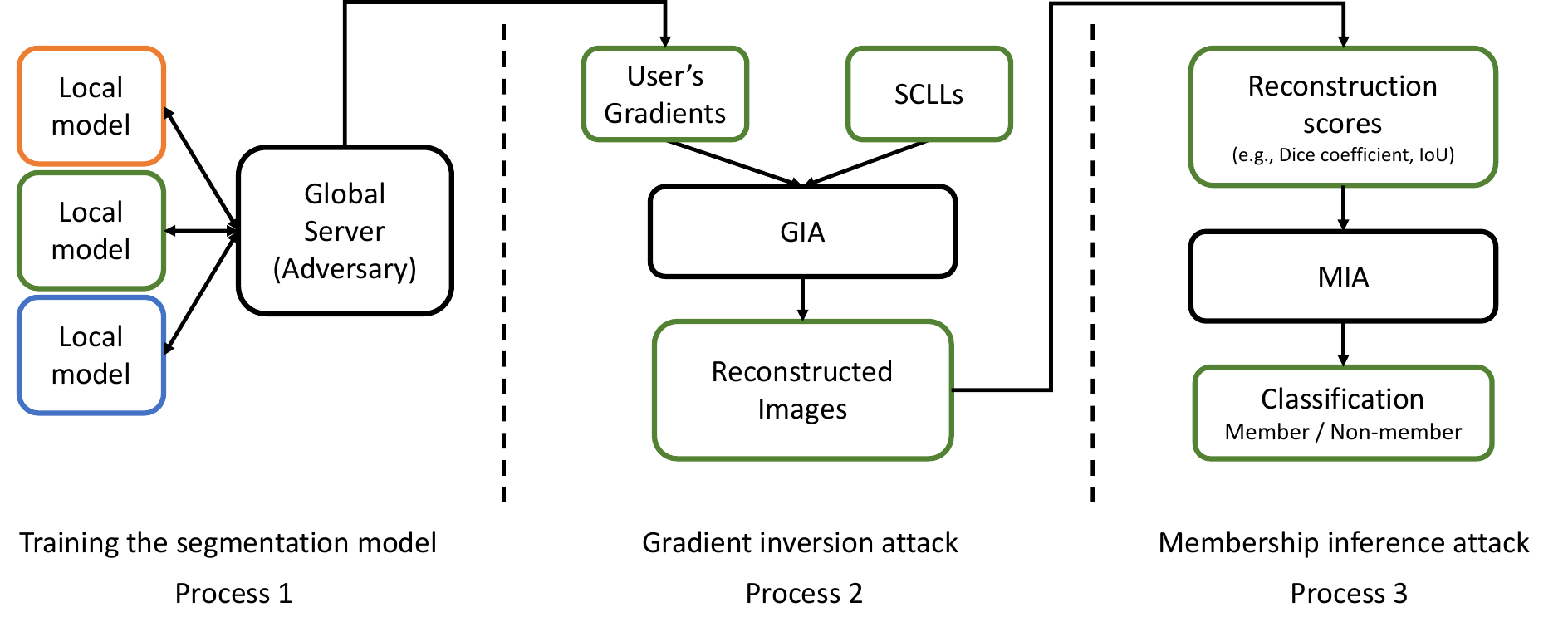}
\vspace{-3mm}
\caption{Overall our proposed MIA method process}
\label{fig:MIA_process}
\vspace{-5mm}
\end{figure*}

\subsubsection{MIA in FL}
Despite blocking access to raw data, FL is vulnerable to MIAs due to the exchanged gradient and weight information. These attacks are broadly categorized into update-based and trend-based methods \cite{bai2024membership}.
\begin{itemize}
    \item Update-based MIAs directly utilize model gradients or parameters. Gradient-based attacks use the original gradients or their differences, whereas parameter-based attacks primarily involve applying the shadow training technique to the FL environment. \cite{nasr2019comprehensive} developed a pioneering MIA that utilizes gradients. While this attack is applicable in various environments, it exhibits two main limitations. First, it assumes the attacker must have access to partial member data from the target client, a condition that is difficult to meet in real-world environments with strong privacy constraints. Second, the process of constructing the attack model is inefficient and time-consuming, as it requires numerous feature vectors from multiple iterations. \cite{melis2019exploiting} exploited the fact that non-zero gradients occurring in the embedding layer expose whether a specific sample participated in training. \cite{li2023effective} likewise proposed passive attack techniques that infer membership information through `gradient-diff' and `cosine' attacks, which calculate the gradient difference between consecutive rounds.
    \item Trend-based MIAs infer membership by tracking changes in specific indicators over the learning process. They perform the attack by analyzing the history of indicators, such as prediction scores or prediction losses, based on the distribution difference between member and non-member data. \cite{gu2022cs} propose the confidence-series-based MIA (CS-MIA), a technique combining advanced confidence metrics with an active attack strategy. To overcome the lack of data needed to build their attack model, the global adversary (the server) actively participates by fine-tuning the global model with auxiliary data and submitting updates as a regular participant. \cite{he2024enhance} proposes a method to strengthen MIAs by utilizing data poisoning and sequential prediction confidence. This method injects malicious data to make the model overlearn information about specific data points, then infers membership by analyzing the changes in confidence vectors over multiple epochs with an AdaBoost classifier. However, because the attack is premised on data poisoning, it has the limitation of being difficult to apply in the `honest-but-curious' adversary scenario, where an attacker only observes model updates.
\end{itemize}

\section{Attack Methodology}\label{sec:methodology}
In this work, we investigate the vulnerability of FL to MIAs \textit{without relying on any auxiliary data from the target (victim)}. Our experimental pipeline, shown in Figure \ref{fig:MIA_process}, proceeds in three stages. First, multiple clients collaboratively train a global segmentation model using standard FL rounds, exchanging only model updates while keeping raw images private. Second, a GIA is performed on the shared model updates to obtain the reconstructed scores. Third, an MIA is then conducted using the reconstructed scores to evaluate the vulnerability of FL to privacy attacks and determine whether sensitive information can be distinguished. This methodology combines a clear definition of the threat model with concrete dataset and environment specifications to enable a systematic analysis of privacy risks in realistic FL deployments.

\subsection{Dataset and Environment Setup}
The experiment uses two distinct datasets of Standard Cell Library Layouts (SCLLs): one comprising metal layers and the other comprising diffusion layers, each from the 32nm and 90nm technology nodes in Synopsys’ Open Educational Design Kit (SAED).
These SCLLs were extracted from the 32nm and 90nm SAED standard cell libraries, using a whitelist to select only the core set of basic logic gates (AND, NAND, OR, NOR, XOR, and XNOR cells) of varying input counts.
In addition to these fundamental gates, any other cell whose associated image dimensions fall at or below the median width and height of the full library is also retained, allowing supplementary standard cells (e.g., inverters, buffers, flip‑flops) that meet the size‑based criterion to be included. This dual‑filter approach ensures the final 141 SCLLs reflect both the essential logic gates and any additional cells that satisfy the dimensional filter.
For the purpose of our segmentation task, we utilized the REFICS\footnote{Link: https://trust-hub.org/\#/data/refics} tool \cite{wilson2021refics} to generate synthetic SEM images and masks, a subset of 141 SEM images with a shot noise parameter set to 20 and 10 $\mu$sec/pixel dwelling time. The background and foreground means are 75 and 135, respectively, with a standard deviation of 20. We resized all images to 256$\times$256 pixels to fit the segmentation model. All experiments were conducted on an AMD EPYC ROME CPU with 32GB of RAM and an NVIDIA B200 GPU with 180GB of GPU RAM. End-to-end, each experiment took around 1 hour to train the FL model and perform GIA and MIA.

\subsection{Threat Model}
The threat model operates under an ``honest-but-curious" assumption, where any participant can become an adversary seeking to uncover sensitive information about other clients via MIAs. Such an adversary is passive, strictly following the FL protocol without tampering with the model or aggregation process. The attack proceeds with the adversary having access to public information such as model weights or gradients, but not private hyperparameters such as the local learning rate $\eta$. In a departure from standard MIA assumptions, this adversary does not need an auxiliary dataset from a similar distribution but instead is equipped with a set of known SCLLs potentially acquired from PDKs.
\vspace{-2mm}
\subsection{Phase 1: Federated Learning}
The experimental setup replicated a FL environment for an image segmentation task. This environment was composed of a central server and two clients, each holding 50 image-mask pairs. The training process employed the FedAvg algorithm, which was deliberately chosen as it represents a more challenging and realistic attack scenario than FedSGD and was conducted over 200 communication rounds, using a segmentation model (U-Net) \cite{ronneberger2015u} as the global model. Within each round, clients completed 2 local training epochs with a batch size of 10 and a learning rate of 0.01.
\vspace{-2mm}
\subsection{Phase 2: GIA with SCLLs}
\subsubsection{Gradient Extraction}
The attack is initiated at a target communication round (e.g., Round 200). The adversary intercepts the model update sent by the target client. The adversary can calculate the client's aggregated gradients ($\nabla \mathcal{L}$) for that training round using the global model weights from the start of the round ($W_{prev}$), the client's updated weights ($W_{curr}$), and an estimated learning rate ($\eta$):
\begin{equation}
\nabla \mathcal{L} = \frac{W_{prev} - W_{curr}}{\eta}
\vspace{-2mm}
\end{equation}
These captured target gradients are a rich source of information about the client's private training batch and serve as the foundational input for the GIA. For our experiments, we set the attacker's estimated learning rate ($\eta$) to 0.01. While our experiments demonstrate the attack's feasibility with a single, assumed learning rate, a more determined adversary could further improve the results. For instance, an attacker could perform a simple grid search over a set of common learning rates and select the one that yields the most plausible reconstructed data.

\begin{figure*}[ht!] 
\vspace{-2mm}
\centering  
\includegraphics[width=0.90\linewidth, height=0.3\linewidth]{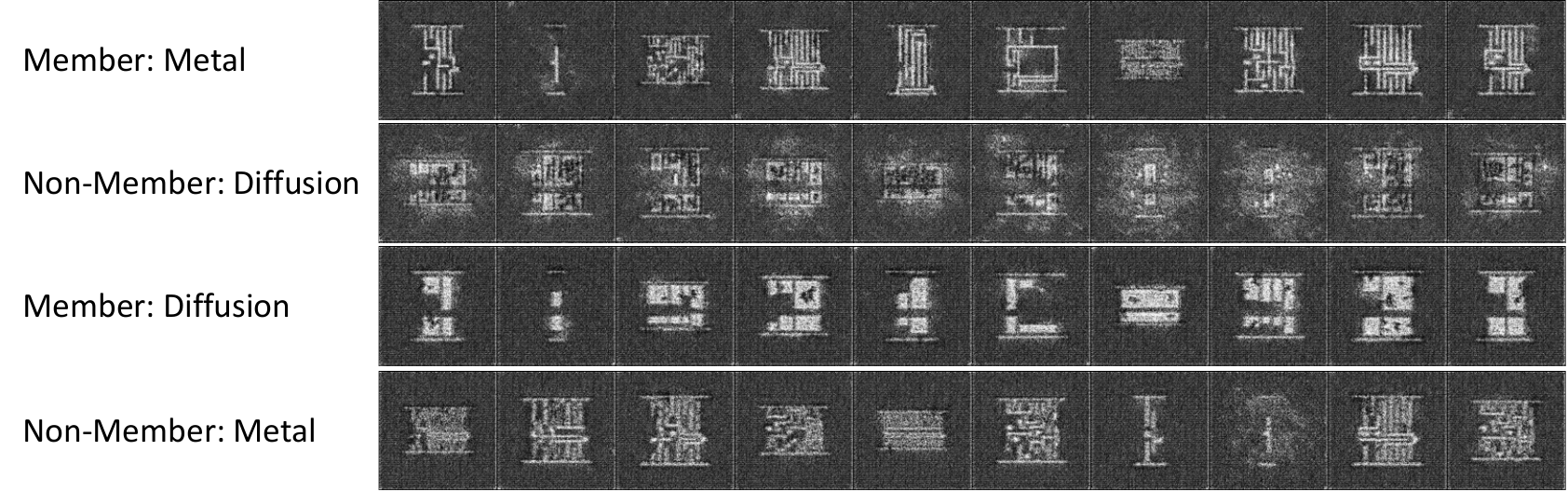} \caption{Representative examples of images reconstructed by GIA, illustrating the large visual quality gap in the metal-as-member case versus the smaller gap in the diffusion-as-member case.} 
\label{fig:reconstructions} 
\vspace{-6mm}
\end{figure*}

\subsubsection{Reconstruction via Guided GIA}
With the target gradients secured, the adversary performs a differential reconstruction to identify the source of the client's data. This process enhances the standard GIA by using SCLLs as a fixed dummy label and comparing the resulting reconstruction quality.
\begin{itemize}
    \item Our key insight is that the quality of a GIA reconstruction can be used to perform a powerful MIA. This is possible because reconstruction quality serves as a direct proxy for the structural correspondence between a client's gradients and a known candidate label. By measuring this correspondence for different candidate labels (e.g., metal vs. diffusion), we can infer membership, as the true member's label will yield a significantly higher reconstruction quality.
    \item The adversary runs the GIA optimization process twice in a blind, comparative manner. In the first run, the GIA is guided by a fixed dummy label ($y'$) from the metal SCLL. This process is then repeated in a second run, this time using a fixed dummy label from the diffusion SCLL. In each run, the optimization begins with a randomly initialized dummy image ($x'$). The optimizer's sole task is to iteratively update (4000 iterations) only the dummy image to minimize the following composite loss function, $L_{total}$:
    \begin{equation}
        L_{total} = L_{grad} + \lambda_{tv} L_{tv} - \lambda_{dummy} L_{dummy}
    \end{equation} 
    The components of this loss function work together to guide the reconstruction. The primary term is the Gradient Matching Loss ($L_{grad}$), which forces the gradients from the dummy data to match the target gradients using a weighted combination of Mean Squared Error (MSE) and Cosine Similarity (CS).
    \begin{align}
        L_{\text{grad}} ={}& \alpha \cdot \text{MSE}(\nabla \mathcal{L}_{dummy}, \nabla \mathcal{L}_{target}) \notag \\
        & + (1 - \alpha) \cdot \text{CS}(\nabla \mathcal{L}_{dummy}, \nabla \mathcal{L}_{target})
    \end{align}
    MSE enforces high pixel-level fidelity, crucial for capturing fine patterns, while CS preserves the global structure and overall shape. The hyperparameter $\alpha$ controls this trade-off: a higher $\alpha$ prioritizes pixel-wise accuracy (MSE), while a lower $\alpha$ emphasizes global structure (CS). Two additional terms complete the loss function. The Total Variation Loss ($L_{tv}$) is added as a regularizer, encouraging the reconstructed image to be smooth by reducing noise and artifacts. Finally, the Forward Pass (FP) Loss ($L_{dummy}$) is the standard forward-pass loss (e.g., MSE) between the model's prediction on the dummy image ($x'$) and the fixed dummy label ($y'$). By subtracting this term, the optimizer is incentivized to find a non-trivial solution that relies heavily on matching the gradients rather than simply fitting the fixed label. For our GIA, we set $\alpha=0.5$, $\lambda_{tv}=0.001$, and the learning rate to 0.01. This dual-run process yields two distinct reconstructed images ($x'_{metal}$ and $x'_{diffusion}$), which are passed to the next phase for analysis.
\end{itemize}
\vspace{-1.5mm}
\subsection{Phase 3: Membership Inference via Mean Threshold Attack} \label{sec:MIA_processing}
The first step is post-processing to quantify the quality of each reconstruction to generate similarity scores. Both reconstructed images from the dual-GIA phase ($x'_{metal}$ and $x'_{diffusion}$) are converted into clean, binary masks. This involves a pipeline of a 5x5 Gaussian blur for denoising, followed by Otsu's automatic thresholding, and finally morphological opening and closing operations to remove noise. Next, a similarity metric, such as the Dice coefficient (Dice), is calculated for each resulting binary mask by comparing it to its corresponding initial SCLL. This process yields two distinct sets of similarity scores: one for the reconstructions guided by the metal SCLL ($S_{metal}$) and another for those guided by the diffusion SCLL ($S_{diffusion}$).
To perform the membership inference, these two sets of scores are then used in a global threshold attack as follows:
\begin{enumerate}
\item \textbf{Score Pooling:} All individual similarity scores from both $S_{metal}$ and $S_{diffusion}$ are combined into a single, comprehensive pool representing all reconstruction quality outcomes.
\item \textbf{Threshold Calculation:} A global classification threshold, T, is established by calculating the statistical mean of this entire pool of scores.
\item \textbf{Classification:} Each individual reconstruction score, regardless of which SCLL was used to guide it, is then compared against this global threshold T. A score higher than T is classified as indicating membership, while a score below T is classified as indicating non-membership.
\item \textbf{Evaluation:} The performance of the MIA is quantified by measuring the success of the final classification step. The evaluation relies on standard metrics, namely Accuracy, Precision, Recall, and the Area Under the Receiver Operating Characteristic Curve (AUC).
\end{enumerate}

\section{Results and Discussion}\label{sec:results}
This section presents the experimental results of our investigation into the vulnerability of FL to MIAs. The findings are organized according to the three-stage experimental pipeline, first presenting the primary attack's performance and then analyzing the contribution of key attack components and the attack's limitations under different data conditions. 

\begin{table}[t] 
\centering 
\caption{Results for Inter-Layer Dice Scores (at $\lambda_{\text{dummy}} = 0$)} 
\label{tab:dice_scores} 
\begin{tabular}{@{}llc@{}} \toprule Target Data & Guiding SCLL & Mean Dice Score \\ \midrule Metal & Metal & \textbf{0.7233} \\ Metal & Diffusion & 0.5876 \\ Diffusion & Diffusion & \textbf{0.8204} \\ Diffusion & Metal & 0.713 \\ \bottomrule 
\end{tabular} 
\vspace{-5mm}
\end{table} 

\subsection{Analysis of Reconstructed Images from GIA} The initial stage of the attack focuses on reconstructing a target client's training data by performing a guided GIA. As per our methodology, for each intercepted gradient, two separate reconstructions were generated: one guided by a fixed dummy label from the metal SCLL ($x'_{\text{metal}}$), and the other by a diffusion SCLL ($x'_{\text{diffusion}}$). A key asymmetry was observed in the quality gap between member and non-member reconstructions. When the target's data was from the metal layer, the visual quality gap in the raw reconstructions was significantly larger: the member ($x'_{\text{metal}}$) reconstruction was high-fidelity while the non-member ($x'_{\text{diffusion}}$) reconstruction was severely degraded. However, when the target's data was from the diffusion layer, the visual quality gap was smaller; both reconstructions were of relatively high quality. Figure \ref{fig:reconstructions} provides representative examples of these outcomes. 
\vspace{-1mm}
\begin{figure}[h!] \centering \begin{subfigure}[b]{0.48\textwidth} 
\vspace{-2mm}
\centering 
\includegraphics[width=\textwidth]{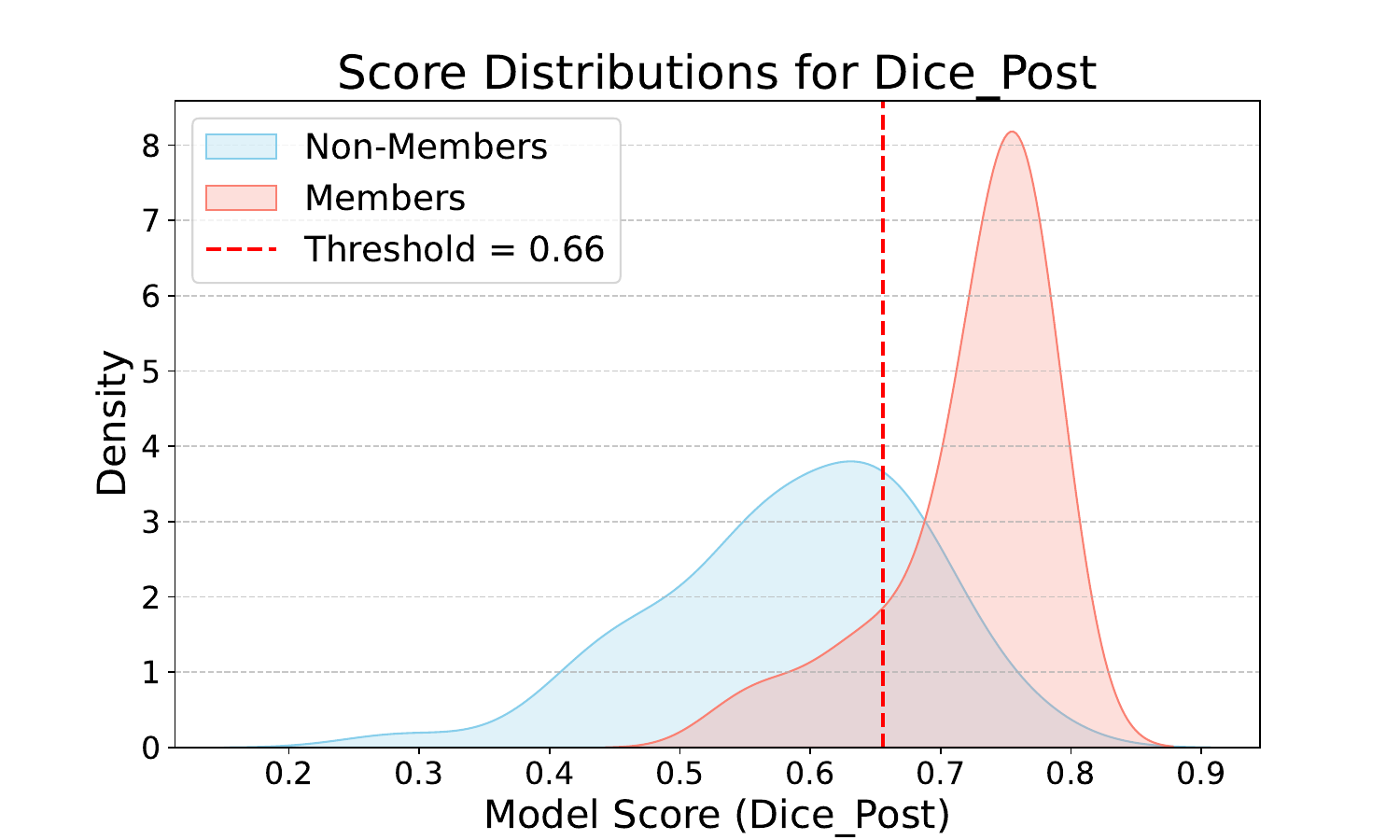} 
\caption{Metal Layer as Member} 
\label{fig:dist_metal} 
\end{subfigure} 
\hfill 
\begin{subfigure}[b]{0.48\textwidth} 
\centering 
\includegraphics[width=\textwidth]{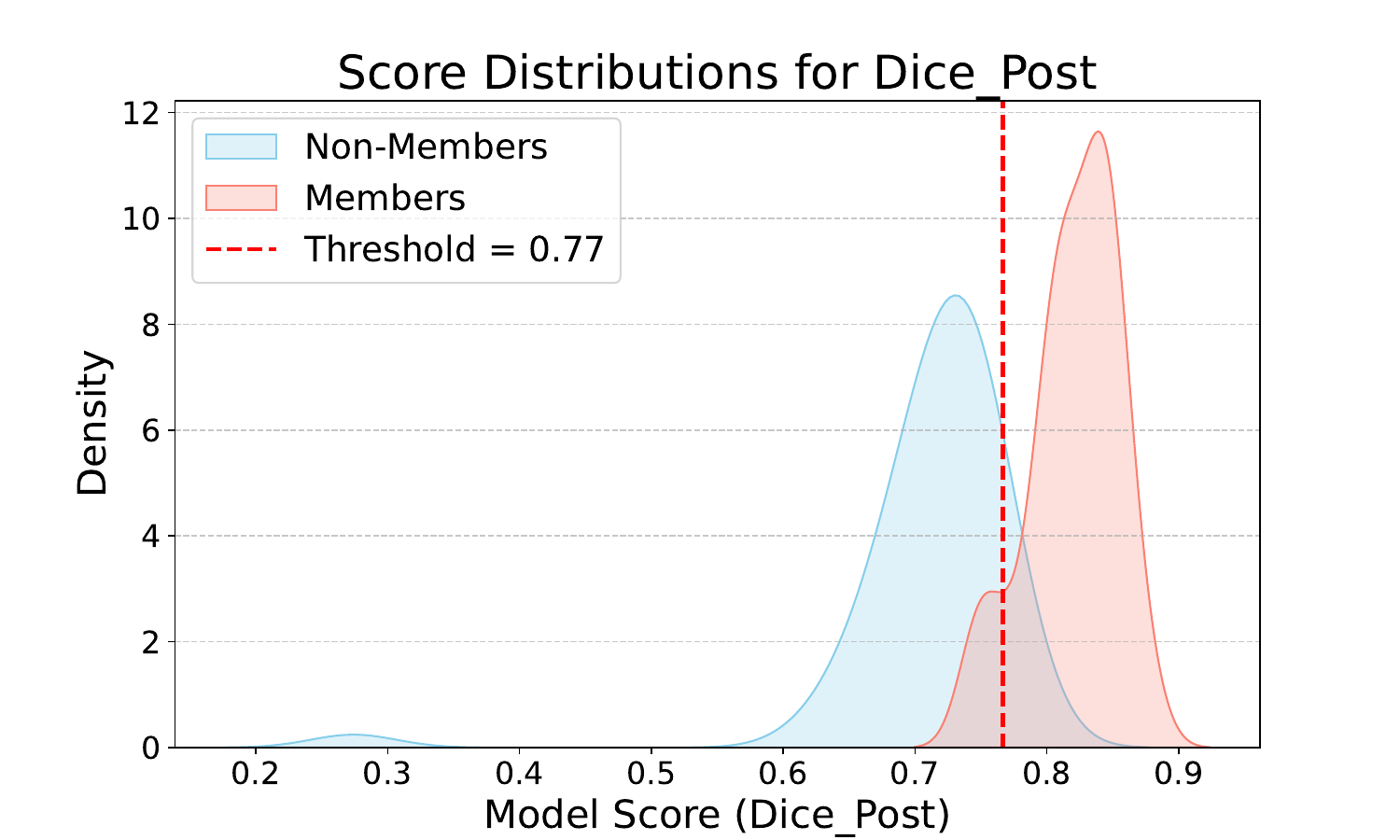} \caption{Diffusion Layer as Member} 
\label{fig:dist_diffusion} 
\end{subfigure} 
\caption{Distribution of Dice scores for members and non-members when $\lambda_{\text{dummy}} = 0$. The significant overlap in (a) contrasts with the clear separation in (b), visually explaining the difference in baseline attack performance.} 
\label{fig:distributions}
\vspace{-4mm}
\end{figure} 

\subsection{Distribution of Reconstruction Similarity Scores} 
\begin{figure}[h!] 
\centering 
\begin{subfigure}[b]{0.2\textwidth} 
\centering 
\includegraphics[width=\textwidth]{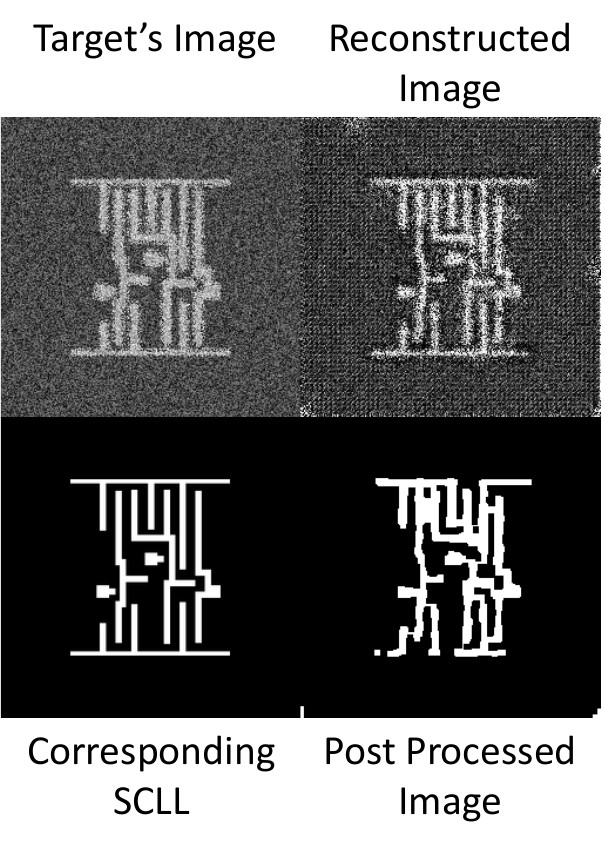} 
\caption{Metal as Member} 
\label{fig:post_metal} 
\end{subfigure} 
\hfill 
\begin{subfigure}[b]{0.2\textwidth} 
\centering 
\includegraphics[width=\textwidth]{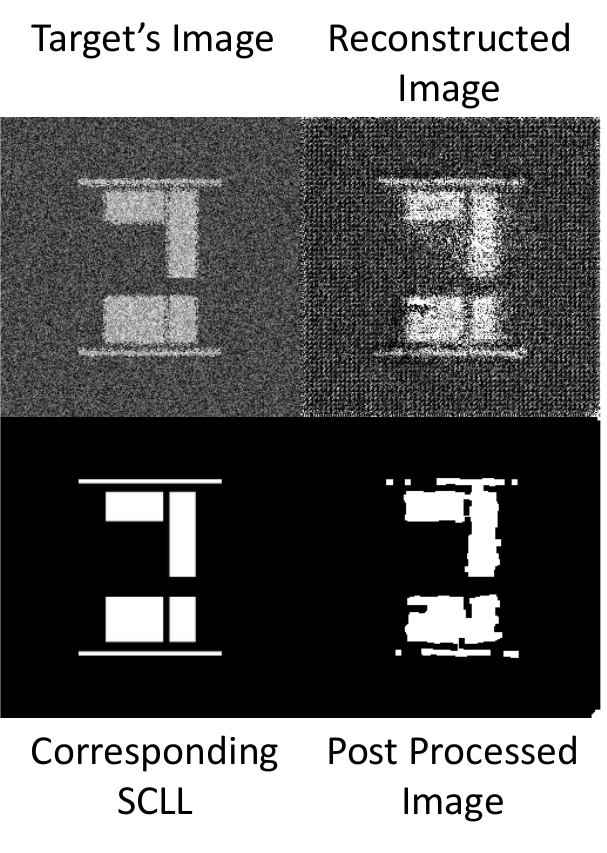} \caption{Diffusion as Member} 
\label{fig:post_diffusion} 
\end{subfigure} 
\caption{Post processed images from each layer.} 
\label{fig:post_processed_examples} 
\vspace{-5mm}
\end{figure}

We generated quantitative similarity scores by converting raw reconstructed images into clean, binary masks (Figure \ref{fig:post_processed_examples}) using the post-processing pipeline from Section~\ref{sec:MIA_processing}.

Following this conversion, we calculated the Dice similarity coefficient for each binary mask. The resulting scores, summarized in Table \ref{tab:dice_scores}, confirm that the quality of the post-processed reconstructions is a strong proxy for membership. As Figure \ref{fig:distributions} shows, the mean Dice score was significantly higher when the reconstruction was guided by an SCLL from the same category as the target's private data.

\begin{figure}[ht] 
\centering  
\includegraphics[width=1\linewidth]{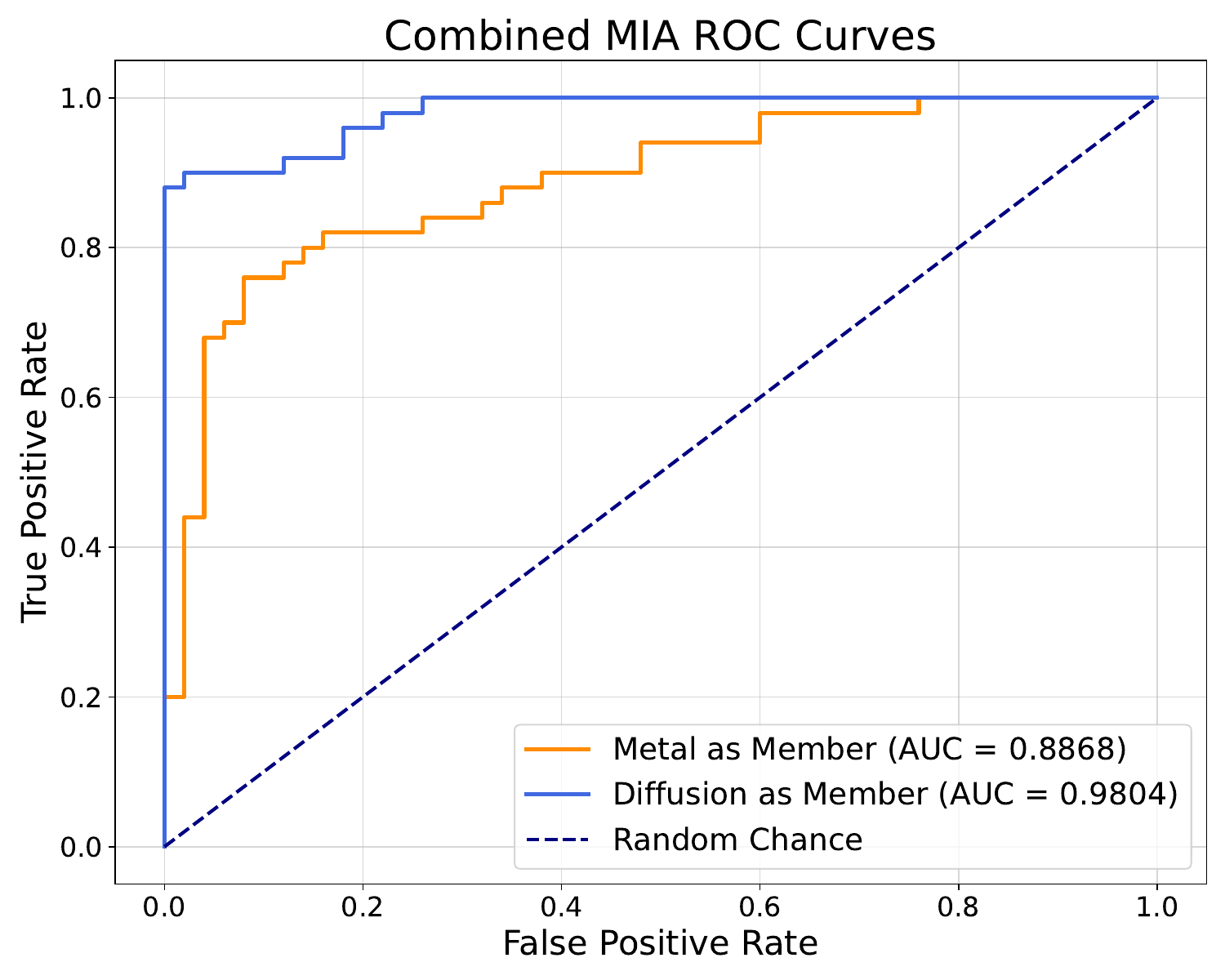} \caption{ROC curves for the membership inference attack on metal vs. diffusion layers, corresponding to the performance metrics in Table \ref{tab:mia_performance}.} 
\label{fig:roc_inter_layer} 
\vspace{-2mm}
\end{figure}

\begin{figure*}[ht!] 
\vspace{-2mm}
\centering  
\includegraphics[width=0.9\linewidth, height=0.3\linewidth]{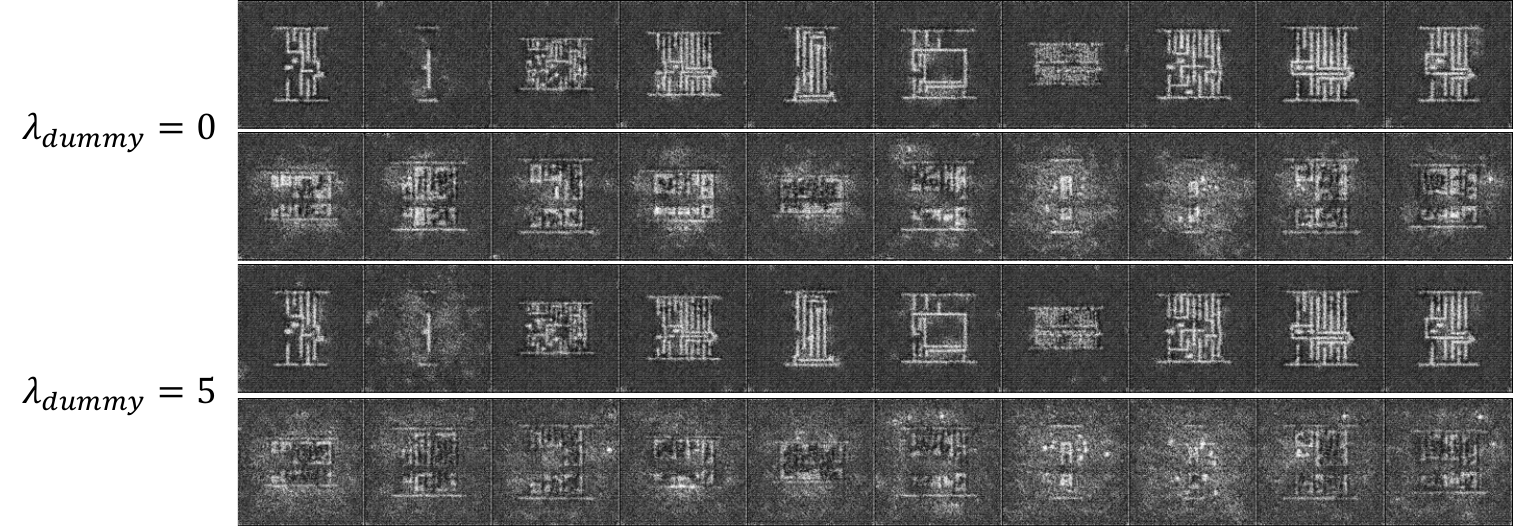} \caption{Visual impact of the $L_{\text{dummy}}$ term when metal is the member. \newline
The quality gap between the member (top row) and non-member (bottom row) reconstructions is visibly larger at the optimal $\lambda_{\text{dummy}}=5$, as the non-member reconstruction becomes significantly more degraded.} 
\label{fig:ablation_reconstructions} 
\vspace{-5mm}
\end{figure*}

\subsection{Membership Inference Attack Performance} 
Using the pooled similarity (Dice) scores from the post-processed images, we performed a membership inference attack. The effectiveness of this attack is detailed in Table \ref{tab:mia_performance} and visualized by the ROC curves in Figure \ref{fig:roc_inter_layer}. While the attack is successful in both scenarios, the results reveal a significant performance asymmetry: the attack on the diffusion layer achieves a near-perfect AUC of 0.9804, whereas the attack on the metal layer yields a substantially lower AUC of 0.8868.

This result is counter-intuitive, as the raw, pre-processed reconstructions (Fig. \ref{fig:reconstructions}) for the metal-as-member case visually exhibit a larger quality gap between member and non-member reconstructions. We attribute this discrepancy to a post-processing bottleneck. The metal layer's fine, complex structures are particularly vulnerable to degradation from our standardized post-processing pipeline (blurring and thresholding). This process can erode the fine lines of a well-reconstructed metal image, artificially lowering its Dice score and making the otherwise clear separation of scores more challenging for a threshold-based attack. In contrast, the simpler, larger shapes of the diffusion layer are more robust to this process, resulting in a cleaner separation of Dice scores and, consequently, a higher AUC. 

\begin{table}[t] 
\centering 
\caption{Metal vs. Diffusion Layers at $\lambda_{\text{dummy}} = 0$} 
\label{tab:mia_performance} 
\begin{tabular}{@{}lcccc@{}} 
\toprule Member Data & Accuracy & Precision & Recall & AUC \\ 
\midrule Metal Layer & 78.00\% & 0.75 & 0.84 & 0.8868 \\
Diffusion Layer & 94.00\% & 0.9783 & 0.90 & 0.9804 \\ 
\bottomrule 
\end{tabular} 
\vspace{-3mm}
\end{table} 

\subsection{Impact of the Forward Pass Loss Term ($L_{\text{dummy}}$)} 
As described in our methodology, the purpose of the negative FP loss term, $L_{\text{dummy}}$, is to serve as a regularizer that forces the optimizer to prioritize matching the target gradients ($L_{\text{grad}}$) over simply fitting the fixed dummy label. Our ablation study was designed to investigate the practical impact of this mechanism, and the results reveal that its importance is directly tied to the challenges of evaluating structurally complex data.
The results, presented in Table \ref{tab:ablation}, show this mechanism is critical for the metal layer. While the standard GIA (`$\lambda_{\text{dummy}} = 0$') produces high-quality raw reconstructions, the post-processing bottleneck degrades their fine-lined structures. This results in a weak measured signal, leading to a low initial AUC of 0.8868. By forcing the optimizer to focus solely on the gradient signal, the primary effect of the $L_{\text{dummy}}$ term is to widen the measurable quality gap. As shown in Figure \ref{fig:ablation_reconstructions}, while the term slightly degrades the quality of the member reconstruction, its effect on the non-member reconstruction is far more severe. This disproportionate degradation creates a more separable distribution of scores after post-processing, boosting the AUC to an optimal 0.9752.

Conversely, for the diffusion layer, whose simpler shapes are robust to post-processing, the gradient signal is already clear and easily measured. The standard GIA is sufficient for a near-perfect AUC of 0.9804 on its own. In this context, forcing the optimizer to focus more intensely on the gradients is an unnecessary constraint and slightly degrades performance. These findings provide compelling evidence that the loss term's primary function is to amplify the true gradient signal when faced with a challenging evaluation scenario.

\begin{table}[t] 
\centering 
\caption{Ablation Study on the FP Loss Term (AUC)} \label{tab:ablation} 
\begin{tabular}{@{}ccc@{}} \toprule $\lambda_{\text{dummy}}$ & Metal Layer (Member) & Diffusion Layer (Member) \\ \midrule \textbf{0 (Term Removed)} & 0.8868 & \textbf{0.9804 (Optimal)} \\ 1 & 0.9376 & 0.9776 \\ 2 & 0.9572 & 0.9684 \\ 3 & 0.9720 & 0.9516 \\ 5 & \textbf{0.9752 (Optimal)} & 0.9228 \\ 10 & 0.9072 & 0.9068 \\ \bottomrule 
\end{tabular} 
\vspace{-4.5mm}
\end{table} 

\subsection{Intra-Layer Attack Performance and Unifying Theory}
We further tested our hypothesis by applying the attack to a more challenging intra-layer scenario: distinguishing between 32nm and 90nm diffusion nodes. This task is inherently more difficult due to the higher structural similarity between two diffusion nodes compared to the clear distinction between metal and diffusion layers. This increased difficulty is reflected in the significantly lower baseline performance (e.g., 0.6688 AUC for the 32nm member) compared to the inter-layer case.

Despite this challenge, the results presented in Table \ref{tab:intra_layer} show a clear parallel to the inter-layer findings and reinforce our theory of a post-processing bottleneck. The 32nm technology node, with its smaller and more complex features, is analogous to the metal layer. Its fine structures are susceptible to degradation from the post-processing pipeline, which explains the poor baseline performance. As with the metal layer, the $L_{\text{dummy}}$ term is essential, creating reconstructions robust enough to survive the evaluation and boosting performance to a near-perfect 0.9916 AUC.

Conversely, the 90nm node, with its larger and structurally simpler features, behaves like the diffusion layer. Its structure is robust to post-processing, allowing the standard GIA to achieve a respectable baseline AUC of 0.8072. In this case, the regularizer is not needed. This confirms that the attack's optimal configuration is not arbitrary, but is directly tied to an interaction between the data's intrinsic structural complexity and the limitations of the evaluation pipeline.

\begin{table}[t]
\centering
\caption{MIA Performance on Intra-Layer Datasets} \label{tab:intra_layer}
\begin{tabular}{@{}lccccc@{}}
\toprule
Member Data & $\lambda_{\text{dummy}}$ & Accuracy & Precision & Recall & AUC \\
\midrule
32nm Diffusion & 0 & 67.00\% & 0.6308 & 0.82 & 0.6688 \\
\textbf{32nm Diffusion} & \textbf{5 (Optimal)} & \textbf{93.00\%} & \textbf{0.8909} & \textbf{0.98} & \textbf{0.9916} \\
\textbf{90nm Diffusion} & \textbf{0 (Optimal)} & \textbf{69.00\%} & \textbf{0.6610} & \textbf{0.78} & \textbf{0.8072} \\
90nm Diffusion & 5 & 64.00\% & 0.6207 & 0.72 & 0.7100 \\
\bottomrule
\end{tabular}
\vspace{-4mm}
\end{table}

\section{Discussion}\label{sec:discussion}
Our work provides a more nuanced understanding of MIAs in FL. We demonstrate that the system's vulnerability is highly conditional, governed by a critical interplay between the intrinsic complexity of private data and the methodology used to evaluate the attack.

Our central finding is a post-processing bottleneck in our evaluation pipeline. We found that the step of converting raw reconstructions into binary masks for scoring disproportionately harms complex, fine-grained data over simpler structures. This bottleneck explains why high-quality raw reconstructions of complex data initially received low attack scores, as the evaluation metric could not accurately capture their fidelity.

This finding reframes our loss term, $L_{\text{dummy}}$, as a mechanism for amplifying the quality gap to overcome the post-processing bottleneck. By forcing the optimization to focus strictly on the gradient, the term degrades the raw quality of both member and non-member reconstructions. Critically, however, this degradation is disproportionately severe for non-members, which lack a true gradient signal to align with. This widens the measurable quality gap between the two classes, creating a more separable distribution of scores and enabling a far more successful attack.
\vspace{-1mm}

\section{Future Works}\label{sec:futurework}
Our findings open several promising avenues for future research, primarily focused on addressing the challenges and extending the capabilities of the attack methodology.

\textbf{1. Mitigating the Post-Processing Bottleneck:} Our identification of the post-processing pipeline as a performance bottleneck is a key finding. Future work should focus on developing more sophisticated evaluation techniques. This could involve creating adaptive post-processing pipelines that adjust their parameters based on the detected complexity of a reconstruction.

\textbf{2. Evaluating Defensive Mechanisms:} Having demonstrated a potent attack, a critical next step is to evaluate potential defenses. Future research should implement and test the robustness of various privacy-preserving techniques against our GIA-based MIA. 

\textbf{3. Automating and Generalizing the Attack:} Our results show that the optimal attack configuration, specifically the `$\lambda_{\text{dummy}}$' hyperparameter, is data-dependent. A valuable line of future work would be to develop methods for automatically tuning this parameter.
This would make the attack more generalizable and effective against a wider range of target data without requiring prior knowledge.

\textbf{4. Scaling the Experimental Setup:} The experiments in this work were conducted in a two-client setting. To better assess the real-world risks, future studies should scale the simulation to include a larger number of clients and investigate the attack's performance in more complex models, thereby stress-testing and validating the threat model.
\vspace{-1mm}

\section{Conclusion}\label{sec:conclusion}
In this paper, we reveal a significant IP risk in FL, challenging its inherent privacy guarantees. We demonstrate that an adversary can infer sensitive IP without auxiliary data, even against a biased evaluation pipeline. Our key insight, gained during the development of our GIA-based MIA, is the discovery of a post-processing bottleneck. We found this flaw, which is specific to reconstruction-based attacks, masks true information leakage by penalizing complex reconstructions. We overcome this measurement flaw with an augmented attack that uses a negative FP loss term ($L_{\text{dummy}}$). This term acts as a conditional amplifier, creating a robust quality gap by disproportionately corrupting non-member reconstructions. Ultimately, this finding poses a severe threat to hardware assurance. By inferring hardware metadata (e.g., technology node or layer), not just individual data, an adversary can accelerate physical attacks like reverse engineering, side-channel attacks, and fault injection.
\vspace{-2mm}

\footnotesize{
\bibliographystyle{IEEEtran}
\bibliography{references}
}

\end{document}